\documentstyle[bezier,preprint,aps]{revtex}
\begin{document}
\tighten
 \title{Exact method for locating  potential resonances
	  and  Regge trajectories}
\author{ S. A.  Sofianos and  S. A. Rakityansky\thanks{Permanent address:
    Joint Institute  for Nuclear Research,Dubna, 141980, Russia}
}
\address{ Physics Department, University of South Africa,  P.O.Box
	  392, Pretoria 0001, South Africa}
\date{\today}
\maketitle
\begin{abstract}
We propose an exact method for locating the zeros of the Jost function
for analytic potentials in the complex momentum--plane. We further  extend
the method to the complex angular--momentum plane  to provide the Regge
trajectories.  It is  shown, by using several examples, that  highly 
accurate results for extremely wide as well as for extremely narrow 
resonances with or without the presence of the Coulomb interaction can be
obtained.
\end{abstract}
\section{Introduction}
Much effort has been devoted in the past to develop  methods to
calculate the energies and widths of resonances in the potential
scattering theory. A comprehensive survey of this subject can be found in
Ref. \cite{kuk}. These methods can be divided into two categories. The
first one is based on methods traditionally employed for real energies
where one  can locate the position of relatively narrow resonances with
a sufficiently high accuracy, but has many difficulties in determining 
their widths and usually  fails for broad resonances. In the second one
the  calculations are performed at complex energies, and therefore the 
widths and  resonant energies are obtained simultaneously.\\

The complex methods have the advantage that the calculations are based
on a rigorous definition of resonances, namely,  as singularities of the
$S$--matrix. Thus, in addition to the information about the resonances 
themselves, they  provide us with information about the analytic
properties of the $S$--matrix and the off--shell properties of the
underlying interaction. However, the existing complex--energy methods
are much more complicated than those of the first group and  require  
sophisticated mathematical and computational methods to handle them.\\

Usually, the complex methods are referred to as the  ``direct calculation
approaches'', but very often with the quotation marks \cite{rescigno}
because
most of them are based on an expansion of the resonant wave function in terms
of square integrable functions and the subsequent determination of
the expansion coefficients either by a diagonalization of the non--Hermitian
Hamiltonian or by a variational procedure.\\

The method we present here formally belongs to the second group i.e., to 
the complex--energy methods. It is based on exact differential equations
for  functions closely related to the Jost solutions and which coincide
with
the Jost functions at large distances \cite{zero,rakpup,nuovocim}. Unlike
the existing  methods, it is simple in applications and although it exploits
the idea of the complex rotation of the coordinates, it is different from 
the traditionally used complex dilation methods in that it does not employ
expansion or variational procedures. Instead,  the Jost  function at a
complex energy is obtained directly from exact equations equivalent to
the Schr\"odinger equation. \\

To demonstrate the effectiveness and accuracy of our approach,
we performed calculations using potentials  previously employed by other
authors \cite{isa,maier,mandel,yamani}. We not only  located the resonances
cited by them but also find  sequences of Jost function zeros corresponding
to broad and extremely narrow resonances which were not considered or missed
in the aforementioned references.\\

In addition to the location of  zeros in the complex $k$--plane, the present
method enables us to locate the zeros of  Jost functions in the complex
angular momentum plane. We demonstrated this by locating  the zeros, 
known as Regge poles, representing resonances in the  complex 
$\ell$--plane.\\

The paper is organized as follows. In Sec. II our formalism  is presented,
while Sec. III is devoted to the boundary conditions. In Sec. IV 
the method is applied to several examples and the results
obtained are discussed and compared with those  obtained by other 
methods.

\section{Basic equations}
Consider the one--channel problem of two, generally charged, particles. 
Appart from the Coulomb force, we assume that these particles interact via 
a central potential $V(r)$ having the properties
\begin{equation}
\label{pot0}
      \lim_{r\to 0} r^2 V(r) = 0
\end{equation}
and
\begin{equation}
\label{potinf}
	\lim_{r\to \infty} rV(r) = 0\,.
\end{equation}
The radial Schr\"odinger equation reads ($\hbar=1$) 
\begin{equation}
\label{schr}
    \left[\partial_r+ 2mk^2 - \ell(\ell+1)/r^2\ - 2\eta k/r\right]
    \Phi_{\ell}(k,r) = V(r)\ \Phi_{\ell}(k,r)\,.
\end{equation}
The regular solutions $\Phi_\ell(k,r)$,  for any complex $k\ne 0$ and complex $\ell$
in the half--plane ${\rm Re\,}\ell>-1/2$, are defined
by the boundary condition
\begin{equation}
\label{bound}
     \lim_{r\to 0}\left [\Phi_\ell(k,r)/F_\ell(\eta,kr)\right ] = 1\,,
\end{equation}
where $F_\ell(\eta,kr)$ is the regular Coulomb function \cite{abram}.\\

In contrast to $\Phi_\ell(k,r)$, the physical solutions (bound, scattering,
and Siegert states) are defined by their behaviour at large distances.
However, they all are regular at $r=0$ and thus  proportional to
$\Phi_\ell(k,r)$. Therefore, once the function $\Phi_\ell(k,r)$  is 
determined at all complex momenta $k$, it represents, in a most general
form, all solutions of physical interest of Eq. (\ref{schr}) since any
physical solution at specific  values of $k$ can be obtained from it 
merely by multiplying $\Phi_\ell(k,r)$ by the proper normalization
constant.\\

At large distances $\Phi_\ell(k,r)$ can be written as a linear combination
\begin{equation}
\label{ass}
   \Phi_\ell(k,r)\ \mathop{\longrightarrow}\limits_{r\to\infty}\ 
   \frac{1}{2}  \left[ H_\ell^{(+)}(\eta,kr) f_\ell^*(\eta,k^*) +
    H_\ell^{(-)}(\eta,kr) f_\ell(\eta,k)\right]\,.
\end{equation}
The functions $H_\ell^{(\pm)}(\eta,z)$ are defined in terms of the
regular $F_\ell(\eta,z)$ and irregular  $G_\ell(\eta,z)$ Coulomb
functions \cite{abram,taylor,gw}),
\begin{equation}
\label{hpm}
      H_\ell^{(\pm)}(\eta,z)\equiv F_\ell(\eta,z)\mp iG_\ell(\eta,z)\,,
\end{equation}
and have the asymptotic behaviour,
\begin{equation}
\label{hass}
      H_\ell^{(\pm)}(\eta,z)\mathop{\longrightarrow}\limits_{|z|\to\infty}
     \mp i \exp\Bigl\{\pm i\left[z-\eta\ln 2z-\ell\pi/2+{\rm arg}
      \Gamma(\ell+1+i\eta)\right]\Bigr\}\,.
\end{equation}
For neutral particles or high energies where the Sommerfeld parameter 
$\eta\rightarrow 0$, the Coulomb functions reduce to the Riccati--Bessel,
Riccati--Neumann, and Riccati--Hankel functions  \cite{abram}, i.e.,
\begin{eqnarray}
\nonumber
 F_\ell(\eta,z) & \mathop{\longrightarrow}\limits_{\eta\to 0} &
				    \jmath_\ell(z)\,,\\
\nonumber
 G_\ell(\eta,z) & \mathop{\longrightarrow}\limits_{\eta\to 0} &
					 -n_\ell(z)\,,\\
\nonumber
 H_\ell^{(\pm)}(\eta,z) & \mathop{\longrightarrow}\limits_{\eta\to 0} &
       h_\ell^{(\pm)}(z)\,.
\end{eqnarray}
The momentum--dependent coefficients in the above linear combination, Eq.
(\ref{ass}), are the Jost functions which are analytical for all complex
$k$ of physical interest and for ${\rm Re\,}\ell>-1/2$. The last restriction
on $\ell$ stems  from the  fact that at ${\rm Re\,}\ell=-1/2$ the role of
$F_\ell(\eta,z)$ and $G_\ell(\eta,z)$,  of being regular and irregular at 
the origin, is interchanged \cite{taylor}.\\

For integer (physical) $\ell$ the Jost function has zeros at a discrete
sequence of points $k_{0i}$, $i=1,2,3,\cdots\ $, situated either on the
imaginary axis of the $k$--plane or under the real axis. At these points 
only the first term  in the asymptotic form (\ref{ass}) survives,
corresponding to either a bound (${\rm Im\, }k>0$) or 
a Siegert (${\rm Im\, }k<0$) state behaviour for large $r$.\\

On the other hand, for real values of $k^2$ (physical energies), the 
function $f_\ell(\eta,k)$ can have zeros at complex $\ell$  which
move, with increasing $k^2$, along the so--called Regge trajectories
which define families of bound and resonant states \cite{taylor}.
Therefore, when the Jost function is known at all complex values 
of $k$ and at all permissible values of $\ell$,  it contains all 
information and characteristics  of the spectrum of the underlying 
physical system.\\

In Ref. \cite{nuovocim} we proposed a method for a direct calculation of
the Jost function for any complex momentum of physical interest. In this 
approach we use a combination of the variable--constant \cite{vcm} and the
complex coordinate--rotation \cite{ccr} methods to solve the Schr\"odinger
equation (\ref{schr}) in an efficient and accurate way without resorting to
any approximation, expansion, or to variational  (stabilization) procedures.
For this, we perform a complex rotation of the coordinate $r$
\begin{equation}
\label{rot}
         r=x \exp(i\theta),\qquad x\ge0,
        \qquad \theta\in[0,\theta_{max}],\quad\theta_{max}<\pi/2
\end{equation}
in the Schr\"odinger equation (\ref{schr}) and look for a solution in the
form
\begin{equation}
\label{ansatz}
      \Phi_\ell(k,r)=\frac12\left[ H_\ell^{(+)}(\eta,kr)
      \ {\cal F}_\ell^{(+)}(\eta,k,x,\theta)+H_\ell^{(-)}(\eta,kr)
     \ {\cal F}_\ell^{(-)}(\eta,k,x,\theta)\right]\,,
\end{equation}
where ${\cal F}_\ell^{(\pm)}(\eta,k,x,\theta)$ are new unknown functions
(variable constants) which at large $x$ become, according to Eq. (\ref{ass}),
true constants. In this way the initial value problem, defined
by Eqs. (\ref{schr}) and  (\ref{bound}), reduces to the following
first--order coupled differential equations
\begin{eqnarray}
     \partial_x\,{\cal F}_\ell^{(+)}(\eta,k,x,\theta) &=&
     \phantom{+}\frac{e^{i\theta}}{2ik}
     H_\ell^{(-)}(\eta,kr)\ V(r)\nonumber\\
  && \times \left[H_\ell^{(+)}(\eta,kr)\ {\cal F}_\ell^{(+)}(\eta,k,x,
     \theta)+
     H_\ell^{(-)}(\eta,kr)\ {\cal F}_\ell^{(-)}(\eta,k,x,\theta)\right]\ ,
\label{eq+}
\end{eqnarray}
\begin{eqnarray}
     \partial_x\,{\cal F}_\ell^{(-)}(\eta,k,x,\theta) &=&
     -\frac{e^{i\theta}}{2ik}
     H_\ell^{(+)}(\eta,kr)\ V(r)\nonumber\\
  && \times \left[H_\ell^{(+)}(\eta,kr)\ {\cal F}_\ell^{(+)}(\eta,k,x,
     \theta)+
     H_\ell^{(-)}(\eta,kr)\ {\cal F}_\ell^{(-)}(\eta,k,x,\theta)\right]\ ,
\label{eq-}
\end{eqnarray}
with the simple boundary conditions
\begin{equation}
\label{cond1}
        {\cal F}_\ell^{(+)}(\eta,k,0,\theta)=
                {\cal F}_\ell^{(-)}(\eta,k,0,\theta)=1\,,
\end{equation}
which follow immediately from Eqs. (\ref{bound}), (\ref{hpm}), and
(\ref{ansatz}).\\

In Ref. \cite{nuovocim} it was proved that if the potential obeys the
conditions (\ref{pot0}) and (\ref{potinf}),  with complex $r$ defined by
(\ref{rot}), then for all complex $k$ situated above the dividing line
$|k|\exp(-i\theta)$  and for  $x\rightarrow\infty$, the function
       ${\cal F}_\ell^{(-)}(\eta,k,x,\theta)$
has a $\theta$--independent limit which coincides with the Jost function,
i.e.,
\begin{equation}
\label{lim}
    \lim_{x\to\infty}{\cal F}_\ell^{(-)}(\eta,k,x,\theta)=f_\ell(\eta,k)\,,
\end{equation}
while the function ${\cal F}_\ell^{(+)}(\eta,k,x,\theta)$ converges to
$f_\ell^*(\eta,k^*)$ at all spectral points $k_{0i}$, $i=1,2,3,\cdots$,
for which $f_\ell(\eta,k_{0i})=0$.\\

The proof was based on the asymptotic behaviour of the functions
$H_\ell^{(\pm)}(\eta,kr)$ at large $r$, and can be generalized to include
the complex angular momentum $\ell$ as well. This generalization is  
straightforward since the angular momentum appears only in the phase factor
of the asymptotic form,  Eq. (\ref{hass}), and therefore for a complex
$\ell$ the functions ${\cal F}_\ell^{(\pm)}(\eta,k,x,\theta)$ at large $x$
have the same behaviour and the asymptotic relation (\ref{lim}) is also 
valid.\\

Thus, the procedure of calculating the Jost function is very simple since
for any fixed pair of $k$ and $\ell$ we only need to solve the system of
first--order differential equations (\ref{eq+}) and (\ref{eq-}) from $x=0$ 
to some $x_{max}$ where ${\cal F}_\ell^{(-)}(\eta,k,x,\theta)$ attains a 
constant value (usually, $x_{max}$ is the range of the potential $V$).
According to Eq. (\ref{lim}), this constant is just the Jost function 
 $f_\ell(\eta,k)$ we are looking for. Simultaneously, as a bonus, we 
obtain the exact wave function in the form of Eq. (\ref{ansatz}) having
the correct asymptotic behaviour.  Depending on the choice of the momentum
$k$, it can be a bound, scattering, or a Siegert state wave function
(rotated when $\theta>0$).\\

The resonances in a specific region of complex $k$ can be easily located 
by taking the rotation angle $\theta$ large enough to cover this region
and then search for zeros of the Jost function. The zeros in the complex
$\ell$--plane can be  similarly located with $\theta=0$.\\

We emphasize that this method is exact. Although  we employ the complex
rotation, we do not need any stabilization procedure. This has been 
demonstrated in Ref. \cite{nuovocim} where we employed an analytically
solvable model  and showed that Eqs. (\ref{eq+}) and  (\ref{eq-}) give
at least 5--digit accuracy for the Jost function in a wide area of complex
$k$  despite the fact that the simplest Runge--Kutta method of integration
was used.

\section{Boundary conditions}
\subsection{Short distances}
Formally, we have to start the integration of Eqs. (\ref{eq+}) and  
(\ref{eq-}) from $x=0$. However, for $\ell\ne 0$ the functions
$H_\ell^{(\pm)}(\eta,kr)$ are irregular, i.e., at the origin
they behave as \cite{messia}
$$
    H_\ell^{(\pm)}(\eta,kr)\ \mathop{\longrightarrow}_{r\to 0}
      \ \frac{\mp i}{2^\ell(2\ell+1)C_\ell(\eta)}
      \left(\frac{kr}{2}\right)^{-\ell}+
     \cases {
      {\cal O}\left(\eta kr\ln kr\right), & for $\ell=0$   \cr
     {\cal O}(\eta(kr)^{1-\ell}), & for $\ell\ne 0$\ ,\cr}
$$
where
$$
      C_\ell(\eta)=\frac{2^\ell \exp(-\pi\eta/2)}{\Gamma(2\ell+2)}
             \left[\Gamma(\ell+1+i\eta)\Gamma(\ell+1-i\eta)\right]^{1/2}\ .
$$
In Ref. \cite{nuovocim} it was shown that the corresponding singularities at
$x=0$ in the above differential equations (\ref{eq+}) and (\ref{eq-})
are integrable when the condition (\ref{pot0}) is fulfilled. Thus,
there is no problem from a formal point of view. However, in
practical calculations  we cannot start  from $x=0$ and therefore 
we have to shift the initial point to some small value $x_{min}$.
Thus, to implement the boundary conditions, we need to know
${\cal F}_\ell^{(\pm)}(\eta,k,x_{min},\theta)$.\\

There are several ways to circumvent this problem. One of them consists in
transforming the differential equations, Eqs. (\ref{eq+}) and  (\ref{eq-}),
into an  equivalent pair of integral Volterra--type equations, viz.,
\begin{equation}
\label{integr}
      {\cal F}_\ell^{(\pm)}(\eta,k,x,\theta)=1\pm\frac{e^{i\theta}}{ik}
      \int_0^xH_\ell^{(\mp)}(\eta,kx'e^{i\theta})V(x'e^{i\theta})
      \Phi_\ell(k,x'e^{i\theta})dx'\ ,
\end{equation}
where $\Phi_\ell(k,r)$ is defined by Eq. (\ref{ansatz}). We can solve these
integral equations in the interval $[0,x_{min}]$ iteratively as 
follows:
\begin{eqnarray}
\label{zero}
       {\cal F}_\ell^{(\pm)\,(0)}(\eta,k,x_{min},\theta) & = & 1\ ,\\
\nonumber
	{\cal F}_\ell^{(\pm)\, (1)}(\eta,k,x_{min},\theta) & = &
          1\pm\frac{e^{i\theta}}{ik}
       \int_0^{x_{min}}H_\ell^{(\mp)}(\eta,kxe^{i\theta})V(xe^{i\theta})
	F_\ell(\eta,kxe^{i\theta})dx\ ,\phantom{*********}\\
\nonumber
&\vdots &\\
\nonumber
{\cal F}_\ell^{(\pm)\, (N)}(\eta,k,x_{min},\theta) & = &
	  1\pm\frac{e^{i\theta}}{2ik}
       \int_0^{x_{min}}H_\ell^{(\mp)}(\eta,kxe^{i\theta})V(xe^{i\theta})
\end{eqnarray}
\begin{eqnarray}
\nonumber
\phantom{********}
\times  \left [ H_\ell^{(+)}(\eta,kxe^{i\theta})
	       {\cal F}^{(+)\,(N-1)}_\ell(\eta,k,x,\theta)
	       +H_\ell^{(-)}(\eta,kxe^{i\theta})
	       {\cal F}^{(-)\,(N-1)}_\ell(\eta,k,x,\theta)\right]\,,
\end{eqnarray}
and then integrate the differential equations starting  from the
value of ${\cal F}_\ell^{(\pm)(N)}(\eta,k,x_{min},\theta)$.\\

For small values of $x_{min}$ the above iteration procedure converges very
fast. Moreover, in implementing the method we found that if ${\rm Re\,}\ell$
is small ($\sim 1$), then a surprisingly high accuracy (better than 
7--digits) can be achieved even with the lowest order iteration, Eq.
(\ref{zero}). For higher values of ${\rm Re\, }\ell$, however, the use
of these simple boundary conditions could result in numerical instabilities.
This is due to the ansatz (\ref{ansatz}) which is  suitable for large
distances, but is not good in the vicinity of  $r=0$. Indeed, near this
point the function $\Phi_\ell(k,r)$, by its definition, is regular and 
therefore the singularities of $H_\ell^{(+)}(\eta,kr)$ and $H_\ell^{(-)}
(\eta,kr)$ are cancelled. This is secured by the boundary conditions 
(\ref{cond1}). In numerical calculations, however, the cancellation of
singularities is always a precarious procedure and a source of possible 
numerical errors. These errors increase with increasing ${\rm Re\, }\ell$
since in this case $H_\ell^{(\pm)}(\eta,kr)$ is more singular.
Therefore, the greater ${\rm Re\, }\ell$ is the further the point $x_{min}$
must be shifted from the origin in order to avoid cancellation errors. This
shift in turn, requires more iterations of Eq. (\ref{integr}) to obtain the
boundary values ${\cal F}_\ell^{(\pm)}(\eta,k,x_{min},\theta)$ to a 
required accuracy.\\

Another way to handle the boundary condition problem is to replace at short
distances the ansatz (\ref{ansatz}) by a  more suitable one. Indeed,
this ansatz was motivated by the variable constant method
\cite{vcm}, i.e. we looked for a solution of Eq. (\ref{schr}) in the form
of a linear combination of its two independent solutions $H_\ell^{(+)}$ and
$H_\ell^{(-)}$ corresponding to $V(r)\equiv 0$.
When the potential is taken into account, the coefficients of this
combination are $r$--dependent and obey the Eqs. (\ref{eq+}) and
(\ref{eq-}). Thus, instead of $H_\ell^{(\pm)}(\eta,kr)$ we can choose
another pair of linearly independent solutions, namely, $F_\ell(\eta,kr)$
and $G_\ell(\eta,kr)$, and the new ansatz reads
\begin{equation}
\label{ansab}
\Phi_\ell(k,r)=F_\ell(\eta,kr)A_\ell(\eta,k,x,\theta)+
	       G_\ell(\eta,kr)B_\ell(\eta,k,x,\theta)\,.
\end{equation}
Since (\ref{ansatz}) and (\ref{ansab}) are merely different representations
of the same function, we have
\begin{equation}
\label{fab}
     {\cal F}_\ell^{(\pm)}(\eta,k,x,\theta) \equiv 
          A_\ell(\eta,k,x,\theta) \pm i B_\ell(\eta,k,x,\theta)\,,
\end{equation}
and the equations for the functions $A_\ell(\eta,k,x,\theta)$ and 
$B_\ell(\eta,k,x,\theta)$ are
\begin{eqnarray}
\nonumber
     \partial_x\,A_\ell(\eta,k,x,\theta) &=&\\
     &\phantom{+}&\frac{e^{i\theta}}{k}G_\ell(\eta,kr)\ V(r)
   \left[F_\ell(\eta,kr)\ A_\ell(\eta,k,x,\theta)+
		  G_\ell(\eta,kr)\ B_\ell(\eta,k,x,\theta)\right]\,,
\label{eqa}\\
\nonumber
     \partial_x\,B_\ell(\eta,k,x,\theta) &=&\\
     &-&\frac{e^{i\theta}}{k}F_\ell(\eta,kr)\ V(r)
   \left[F_\ell(\eta,kr)\ A_\ell(\eta,k,x,\theta)+
		  G_\ell(\eta,kr)\ B_\ell(\eta,k,x,\theta)\right]\,.
\label{eqb}
\end{eqnarray}
The corresponding boundary conditions,
\begin{equation}
\label{condab}
        A_\ell(\eta,k,0,\theta)=1\,,\qquad 
       B_\ell(\eta,k,0,\theta)=0\,,
\end{equation}
follow immediately from (\ref{cond1}) and (\ref{fab}).\\

In other words, we have two equivalent pairs of equations, Eqs. (\ref{eq+})
and (\ref{eq-}) and   (\ref{eqa}) and (\ref{eqb}), defining the same function
$\Phi_\ell(k,r)$ in its two different representations (\ref{ansatz}) and
 (\ref{ansab}).  Computationally it is easier to start  the integration 
of equations (\ref{eqa}) and (\ref{eqb}) at $x_{min}$ and continue it
up to some intermediate point $x_{int}$ (not necessary small), and then
to integrate the equations (\ref{eq+}) and  (\ref{eq-}) from $x_{int}$
to $x_{max}$ where ${\cal F}_\ell^{(-)}(\eta,k,x_{max},\theta)$ coincides
with the Jost function.\\

Similarly to equations (\ref{eq+}) and (\ref{eq-}) the differential equations
for $A_\ell(\eta,k,x,\theta)$ and $B_\ell(\eta,k,x,\theta)$  can  be
transformed into integral Volterra--type equations,
\begin{eqnarray}
\label{integra}
     A_\ell(\eta,k,x,\theta)&=&1+\frac{e^{i\theta}}{k}
     \int_0^xG_\ell(\eta,kx'e^{i\theta})V(x'e^{i\theta})
     \Phi_\ell(k,x'e^{i\theta})dx'\,,\\
\label{integrb}
    B_\ell(\eta,k,x,\theta)&=&\phantom{1}-\frac{e^{i\theta}}{k}
     \int_0^xF_\ell(\eta,kx'e^{i\theta})V(x'e^{i\theta})
     \Phi_\ell(k,x'e^{i\theta})dx'\,,
\end{eqnarray}
where $\Phi_\ell(k,r)$ is defined by Eq. (\ref{ansab}).  Iterations of
these integral equations can also be used to obtain corrections, if 
necessary,  to the simplest form of the boundary conditions, namely,
\begin{equation}
\label{condabx}
  A_\ell(\eta,k,x_{min},\theta)=1\ ,\quad B_\ell(\eta,k,x_{min},\theta)=0\,.
\end{equation}
\subsection{Large distances}
The behaviour of $\Phi_\ell(k,r)$ at large distance, is determined by the
functions $H_\ell^{(\pm)}(\eta,kr)$. Therefore, the correct 
asymptotic form  is  automatically secured.\\

Indeed, suppose we have found on the positive imaginary axis of the
 $k$--plane a value $k_0$ for which  
           ${\cal F}_\ell^{(-)}(\eta,k_0,x_{max},0)=0$
(when ${\rm Im\, }k\ge 0$ we can always put $\theta=0$), i.e., 
we located a zero of the Jost function corresponding to a bound state. 
The physical bound state wave function is then given by,
$$
    \varphi_\ell^{bound}( k_0,r) = {\cal N}\Phi_\ell(k_0,r)\, ,
$$
and differs from $\Phi_\ell(k_0,r)$ only by a normalization 
factor ${\cal N}$ which can, in principle, be
found along with $\Phi_\ell(k_0,r)$ in terms of the Jost function and its
derivative \cite{gw,sitenko}, or simply  from the normalization integral.
At large $r$ only the first term of Eq. (\ref{ansatz}) survives, i.e.,
\begin{equation}
\label{bass}
    \Phi_\ell(k_0,r)\ \mathop{\longrightarrow}_{|r|\to\infty}\ \frac12
    {\cal F}_\ell^{(+)}(\eta,k_0,x_{max},0)H_\ell^{(+)}(\eta,k_0r)\,.
\end{equation}
Obviously, in this expression the exponentially decaying tail of the bound
state wave function is presented in an exact form.\\

For scattering states (real positive $k$), the asymptotic
form of $\Phi_\ell(k,r)$ contains both terms of Eq. (\ref{ansatz}),
\begin{equation}
\label{scass}
    \Phi_\ell(k,r)\ \mathop{\longrightarrow}_{|r|\to\infty}\ \frac12\left[
     H_\ell^{(+)}(\eta,kr)
    \ {\cal F}_\ell^{(+)}(\eta,k,x_{max},0)+H_\ell^{(-)}(\eta,kr)
     \ {\cal F}_\ell^{(-)}(\eta,k,x_{max},0)\right]\,,
\end{equation}
where the functions $H_\ell^{(\pm)}(\eta,kr)$ represent the incoming and
outgoing spherical waves (again in the exact form). The scattering wave
function, $\varphi_{\ell,\, k}^{scatt}(r)$, differs from $\Phi_\ell(k,r)$
only by a constant factor, viz.,
\begin{equation}
\label{scat}
     \varphi_{\ell,\, k}^{scatt}(r)\ =\ \frac{1}{2{\cal F}_\ell^{
      (-)}(\eta,k,x_{max},0)}\left[ H_\ell^{(+)}(\eta,kr)
     \ {\cal F}_\ell^{(+)}(\eta,k,x,0)+H_\ell^{(-)}(\eta,kr)
     \ {\cal F}_\ell^{(-)}(\eta,k,x,0)\right]\,,
\end{equation}
where we assumed that  the scattering states $|\Psi_{\vec k}^{scatt}\rangle$
are normalized according to
$$
    \langle\Psi_{\vec k'}^{scatt}|\Psi_{\vec k}^{scatt}\rangle=\delta
(\vec {k'}-\vec k)
$$
and expanded in partial waves as follows:
$$
    \langle\vec r|\Psi_{\vec k}^{scatt}\rangle\ =\ \sqrt{\frac{2}{\pi}}
\frac{1}{kr}\mathop{\sum}_{\ell m}i^\ell\varphi_{\ell,\, k}^{scatt}(r)
Y_{\ell m}^*(\hat{\vec k})Y_{\ell m}(\hat{\vec r})\,.
$$

For the Siegert states, corresponding  to zeros of
${\cal F}_\ell^{(-)}(\eta,k,x_{max},\theta)$ in the lower half of the
$k$--plane, we have the same kind of asymptotic behaviour as in
$\Phi_\ell(k_0,r)$ given by Eq. (\ref{bass}) but in this case $\theta>0$,
and therefore such states can be treated and normalized similarly to bound
states \cite{rescigno}.\\

In summary, the representation (\ref{ansab}) secures the proper behaviour
of $\Phi_\ell(k,r)$ at short distances, while the representation
(\ref{ansatz}) guarantees its correct behaviour at large $r$. The use of
these representations enables us to achieve high accuracy in the 
solution of Eq. (\ref{schr}) at all complex values of $k$.\\

\section{Numerical examples}
In order to demonstrate the ability and accuracy of the proposed method, we
chose two simple potentials previously used in Refs. 
\cite{isa,maier,mandel,yamani}. This choice is further motivated by the
richness of the spectra  generated by these potentials and 
by their simple form. And, as we found, their spectra include very wide as
well as extremely narrow resonances which are difficult  to locate  with
most of the existing methods. Thus they are ideally suited as testing
cases.\\

In atomic units \cite{landau}, these potentials have
the following form
$$
     V_1(r)=7.5r^2\exp(-r)+\frac{z}{r}
$$
and
$$
     V_2(r)=5\exp\left[-0.25(r-3.5)^2\right]-8\exp\left(-0.2r^2\right)\,.
$$
The reader not accustomed to the atomic units, may  assume that
the above potentials are given in MeV and the distances in 
fm. In such a  case  $\hbar^2/2m=1/2$ MeV\,fm$^2$  while the 
 Sommerfeld  parameter is given by $\eta=z/k$. Then the numerical values
of the resonance energies are the same and independent of the unit used
(MeV or atomic units a.u.). In what follows, in order to avoid
possible misunderstanding, we will use the  MeV-fm units. We note that
$V_1(r)$ is a good case to test the ability of  the  method to deal with
interactions having a Coulomb tail. Similarly to Ref. \cite{yamani} we  
assumed that the Coulomb part is attractive with  $z=-1$. In order to
compare our results with those given in Refs. \cite{isa,maier,mandel},
we also  performed calculations  with $z=0$.\\

To locate zeros of the Jost function in the complex
$k$--plane as well as in
the complex $\ell$--plane, we searched for minima of its modulus,
$|{\cal F}_\ell^{(-)}(\eta,k,x_{max},\theta)|$, considered as a function of
two variables, either ${\rm Re\,}k$ and ${\rm Im\,}k$ or ${\rm Re\,}\ell$
and ${\rm Im\,}\ell$. This is based on the so called maximum modulus
principle for a complex--valued function \cite{encikl}. According to this
principle, when a function $f(z)$ is holomorphic and not constant in a
region $D$ of the complex plane, $|f(z)|$ can never attain its maximum
in the interior of $D$ but only on the boundary of $D$. Therefore
the minima of $|f(z)|$ must coincide with the zeros of $f(z)$. Indeed,
assuming that $|f(z)|$ has a minimum at $z=z_0$ inside the area $D$ and
$f(z_0)\ne 0$ then around the point $z_0$ the function $1/f(z)$ is
holomorphic and has a maximum inside an area, which contradicts the
maximum modulus principle. Thus, if  a minimum of $|f(z)|$ is found, it
must be the  zero of $f(z)$.\\

The located zeros of the Jost function  in the fourth quadrant of the
complex $k$--plane  for the potential $V_1(r)$  for the $\ell=0$ partial
wave, and the corresponding resonance energies and widths  are presented in
Table I together with the results of Refs. \cite{isa,maier,mandel,yamani}.
Our  calculations  have been performed with the simplest boundary  condition
(\ref{condabx}). This was sufficient  to achieve an accuracy of at least 9
digits. This is checked by changing the rotation angle $\theta$ since
the Jost function must be $\theta$--independent. We note that only  few 
$S$--wave resonances are presented although many more were located. The 
reason is that this potential has been used earlier by  several authors (but
only for the case $\ell=0$) and thus we employed it in order to compare our
results with those of Refs.  \cite{isa,maier,mandel,yamani}.  The 
sequence of the $S$--wave resonant zeros continues downwards in the complex
$k$--plane. This behaviour can be  seen in Fig. 1 where these zeros are
plotted. A similar behaviour was also found for resonances of higher partial
waves.  In these cases, however, the first few zeros of the sequences are
closer to the real axis (which  means that they have a smaller width). The
same  trends were found for the resonances of the  potential $V_2(r)$ which
generates a richer spectrum.\\

Comparison with the other calculations for the potential $V_1(r)$ shows 
(see Table I) that only the  ``direct'' (dilatation) method of Ref.
\cite{isa} gives an accuracy which is comparable with ours. In that
reference, the 5--digit accuracy was achieved by using over 40 
exponentially decreasing functions in the  expansion of the rotated 
Siegert state. In contrast a 9--digits accuracy is achieved by our
method without  any exertion and, if neccessary, can easily be improved.
Such an improvement is of crucial importance when one deals with extremely 
narrow resonances. When the width of a resonance is 7 orders of
magnitude less than its energy, one needs at least a 7--digit accuracy 
to be able to discern it.\\

Another extreme situation  is the case of  very broad resonances, i.e., when
the Jost function zeros are situated far below the real axis. As can be seen
in Table I, even for the first resonance ($E_0=3.426390331$ MeV,
$\Gamma=0.025548962$ MeV) only a third digit accuracy in the width
has been achieved by  the ``real--energy methods'' of Refs. 
\cite{maier,mandel} and by the ``semi-complex method'' of Ref. \cite{yamani}
while the next, moderately broad,  resonance ($E_0=4.834806841$ MeV, 
$\Gamma=2.235753338$ MeV), is already beyond their resolution.\\

The ``real--energy methods'' of Refs. \cite{maier,mandel}  consider
eigenenergies of a system enclosed in a box. These eigenenergies are moving
together with the change of the radius of the  box, generating the so--called
quasi--crossings at resonance energies. The width of a resonance is
determined by the breadth of the quasi--crossing. Such an approach exploits
the physical idea that the resonant states, are only slightly affected  by
variations  of the  radius of the box, in contrast to the spurious states 
that emerge in the box.  It seems, however, that this idea is not suitable 
for broad resonant states. Indeed,  in Fig. 1 of Ref. \cite{maier} and
Fig. 1  of Ref. \cite{mandel}, presenting the box eigenenergies,  no traces
of the the second  quasi--crossing of the $S$--wave resonance at 
$E=4.834806841$ MeV  or of the third at $E=5.277279780$ MeV  can be found.\\

It is claimed that broad resonances are unimportant and thus the 
inability of a method to describe them is a minor drawback. However, in
certain  physical systems, broad resonances play a significant role. An
example is the $S_{11}(1535)$ resonance of the interaction of
the $\eta$--meson with a nucleon which lies 48 MeV above the threshold
while its width is 150 MeV \cite{PDGr}. Nevertheless it prescribes the
dynamics of the $\eta$--meson interaction with nuclei. In particular,
due to this  resonanance, certain $\eta$--nucleus systems can 
have quasi--bound states \cite{PRRapidCom}.\\

One of the advantages of the exact method presented in this work,
is that bound, scattering, and
resonant states can be treated in a uniform way regardless of their width.
All spectral points  can be located with the same accuracy irrespective
of their location on the complex $k$--plane.  This is exemplified by the
spectral analysis of the potential $V_2(r)$. Sequences of the bound and 
resonant states generated by this potential are given in Tables II and III.
The $S$-- and $P$--wave sequences are also shown in Fig. 2.\\

The $S$--wave states, generated by the potential $V_2(r)$, were previously
considered in Ref. \cite{maier}, where only the first two resonances were
found. As can be seen in Table II, the spherical--box method of Ref.
\cite{maier}  provides the position of the first (narrow) resonance fairly 
well. However, the width is obtained only to two correct digits. For the
second resonance, which is broader than the first one, the accuracy 
for the energy evaluation is down to 3 digits and for the width just to 1
 digit. The other resonances of the $S$--wave sequence, 
presented in Table II, were not obtained with the box--method.\\

The accuracy of the present  method is determined by the accuracy of solution
of the differential equations. By choosing  the tollerance to be small enough
we can locate practically all resonances.  Two remarkable examples of 
extremely narrow  resonances, generated by the potential $V_2(r)$, are the
one found for the $P$--wave
at $E_0=0.807634844$ MeV  ($\Gamma=0.110\times 10^{-6}$ MeV) and the other
for  the $F$--wave at $E_0=1.009031953$ MeV ($\Gamma=0.46\times 10^{-7}$ 
MeV) and are  parts of the  sequences presented in Table III. 
If the attraction of the potential is slightly increased, these extremely
narrow resonances become bound states in the $P$ and $F$ partial waves.\\

Some of the Jost--function zeros found in the complex $\ell$--plane are given
in Tables IV and V. The corresponding Regge trajectories are depicted in
Fig. 3. Each trajectory begins from an $S$--wave spectral state.
The first one begins
from the ground state and passes via the lowest states of each  partial wave.
When the  energy is negative, the trajectory lies on the real axis, while
at positive energies it gradually goes upwards. We note that the width of
a narrow resonance can, in principle, be found via ${\rm Im\,}\ell$ which
corresponds to  an integer value of ${\rm Re\,}\ell$. However, the relation
between $\Gamma$ and ${\rm Im\,}\ell$ involves also the 
derivative of the trajectory \cite{sitenko}.\\

The Regge trajectories, can provide us with a useful information on the
spectrum of the physical system. They combine bound and resonant
states in families and therefore in calculating such trajectories, we can
find out at which energies and in which partial waves resonances must exist.
Furthermore, the Regge trajectories give us a general insight into the 
spectrum of the Hamiltonian under consideration. For example, we can state
that the $F$--wave resonance at $E=1.009031953$ MeV is the narrowest one
for the potential $V_2(r)$. This follows from the fact that this resonance
belongs to the lowest Regge trajectory and is the first on it.\\

In conclusion, the method we present in this article, enables us to locate
the zeros of the Jost function in the complex  $k$--plane and to calculate
the Regge trajectories  in the complex $\ell$--plane in a simple, efficient,
and extremely accurate way. To the best of our knowledge, no other method
achieved such a performance in the past. Since it is based on
exact equations, the method simultaneously provide us with the corresponding
wave functions of the spectral states.\\

As a final note we mention  that the method can be easily generalized to
treat the coupled channel problems having the same angular momentum. In such 
a case we replace the potential and the functions 
${\cal F}_\ell^{(\pm)}(\eta,k,x,\theta)$ by  matrices and the spectrum is 
then defined by the zeros of the Jost--matrix determinant. Channels of
different angular momenta can also be treated in the same manner, but 
this requires a more elaborate treatment of the boundary conditions at 
$x=0$ since some off--diagonal elements of the matrices 
${\cal F}^{(\pm)}_\ell$ for different $\ell$ may be singular at the origin.
Nonlocal potentials can also be considered within the proposed approach. 
Work on all these generalizations is under way.

\bigskip
\acknowledgements
{One  of us (S.A.R)  gratefully acknowledges  financial support
from the University of South Africa and
the Joint Institute  for Nuclear Research, Dubna.}

\newpage

%
\begin{table}
\caption[]{The zeros $k_0$ of the Jost function
in the complex $k$--plane and the energies $E_0$ and widths $\Gamma$
of the $S$--wave resonances for the potential $V_1(r)$ with $(z=-1)$ and
 without $(z=0)$ the Coulomb term.}
\begin{tabular}{|r|r|r|r|c|}
\hline
 $z$\ & $ k_0\ ({\rm fm}^{-1})$\phantom{0000000} &
 $E_0\ ({\rm MeV}) $\phantom{000} &
 $\Gamma\ ({\rm MeV})^{\mathstrut}_{\mathstrut} $\phantom{000} & Ref.\\
\hline
0\ & $2.617786172-i0.004879876 $ & 3.426390331 & 0.025548962 & this work\\
   & & 3.42639\phantom{0000}  & 0.025549\phantom{000}    & \cite{isa}   \\
   & & 3.4257\phantom{00000}  & 0.0256\phantom{00000}    & \cite{maier} \\
   & & 3.426\phantom{000000}  & 0.0256\phantom{00000}    & \cite{mandel}\\
   & & 3.426\phantom{000000}  & 0.0258\phantom{00000}    & \cite{yamani}\\
0\ & $3.130042444-i0.357144253 $ & 4.834806841 & 2.235753338 & this work\\
0\ & $3.398392393-i0.997251873 $ & 5.277279780 & 6.778106356 & this work\\
\hline
-1\ &$1.887074210-i0.000025362 $ & 1.780524536 & 0.000095719 & this work\\
    &   & 1.7805\phantom{00000} & 0.0000958\phantom{00} & \cite{yamani}\\
-1\ &$2.871167766-i0.201530270 $ & 4.101494946 & 1.157254428 & this work\\
-1\ &$3.169186525-i0.846652839 $ & 4.663461099 & 5.366401539 & this work\\
\hline
\end{tabular}
\end{table}
%
\begin{table}
\caption[]{The zeros  $k_0$ of the Jost function 
in the complex $k$--plane and the energies $E_0$ and widths $\Gamma$
 of the $S$--wave bound and resonant states  for the potential $V_2(r)$.}
\begin{tabular}{|c|r|r|r|c|}
\hline
 $\ \ell\ $ & $ k_0\ ({\rm fm}^{-1})$\phantom{--------} &
 $E_0\ ({\rm MeV})\phantom{--} $ &
 $\Gamma\ ({\rm MeV})^{\mathstrut}_{\mathstrut}\phantom{--} $ & Ref.\\
\hline
0 & $\phantom{0.000000000+}i3.023634507 $ & -4.571182814 & 0  & this work\\
0 & $\phantom{0.000000000+}i1.329872758 $ & -0.884280776 & 0  & this work\\
0 & $2.122442334-i0.000027859 $ & 2.252380731 & 0.000118256 & this work\\
  &    & 2.25237\phantom{0000} & 0.0001196\phantom{00} &   \cite{maier}\\
0 & $3.000600515-i0.041316851 $ & 4.500948186 & 0.247950731 & this work\\
  &  & 4.50\phantom{0000000} & 0.28\phantom{0000000}   & \cite{maier}\\
0 & $3.485234669-i0.360967988 $ & 6.008281406 & 2.516116297 & this work\\
0 & $3.974580284-i0.788297975 $ & 7.587937367 & 6.266307179 & this work\\
0 & $4.454733926-i1.227097054 $ & 9.169443586 &10.932781876 & this work\\
0 & $4.922800349-i1.664647328 $ &10.731456273 &16.389452891 & this work\\
0 & $5.378677040-i2.097566794 $ &12.265190122 &22.564268707 & this work\\
\hline
\end{tabular}
\end{table}
%
\begin{table}
\caption[]{The zeros  $k_0$ of the Jost function 
in the complex $k$--plane and the energies $E_0$ and widths $\Gamma$
of the bound and resonant states  for the potential $V_2(r)\,$ in several
higher  partial waves $(1\le\ell\le 4)\,$.}
\begin{tabular}{|c|r|r|r|}
\hline
 $\ \ell\ $ & $ k_0\ ({\rm fm}^{-1})$\phantom{-----} &
 $E_0\ ({\rm MeV}) $\phantom{--} &
 $\Gamma\ ({\rm MeV})^{\mathstrut}_{\mathstrut} $\phantom{--} \\
\hline
1 & $ i2.289054013 $ & -2.619884138 & 0\phantom{.000000000} \\
1 & $1.270932606-i0.000000043 $ & 0.807634844 & 0.000000110 \\
1 & $2.674841953-i0.002433228 $ & 3.577386775 & 0.013017001 \\
1 & $3.263588553-i0.162882441 $ & 5.312239776 & 1.063162540 \\
1 & $3.742982846-i0.564324132 $ & 6.845729429 & 4.224511090 \\
1 & $4.225164261-i0.999000737 $ & 8.427005280 & 8.441884426 \\
1 & $4.696779393-i1.437816345 $ & 9.996210410 &13.506212364 \\
1 & $5.156711217-i1.873567558 $ &11.540707589 &19.322893683 \\
\hline
2 & $ i1.232503483 $ & -0.759532418 & 0\phantom{.000000000} \\
2 & $2.183644493-i0.000018973 $ & 2.384151637 & 0.000082862 \\
2 & $3.052966547-i0.035600651 $ & 4.659668666 & 0.217375191 \\
2 & $3.527492840-i0.341406094 $ & 6.163323809 & 2.408615103 \\
\hline
3 & $1.420585762-i0.000000016 $ & 1.009031953 & 0.000000046 \\
3 & $2.760769155-i0.001484354 $ & 3.810922062 & 0.008195917 \\
3 & $3.343986231-i0.139396664 $ & 5.581406240 & 0.932281051 \\
3 & $3.802701793-i0.526397919 $ & 7.091723079 & 4.003468620 \\
\hline
4 & $2.313665822-i0.000006013 $ & 2.676524768 & 0.000027824 \\
4 & $3.170315114-i0.023354979 $ & 5.025176235 & 0.148085287 \\
4 & $3.623738918-i0.294896551 $ & 6.522259887 & 2.137256217 \\
4 & $4.076486137-i0.706384524 $ & 8.059380065 & 5.759133441 \\
\hline
\end{tabular}
\end{table}
%
\begin{table}
\caption[]{Two lowest Regge trajectories for the potential $V_2(r)\,$.
Only those points  which correspond to the bound and resonant
energies are given.}
\begin{tabular}{|r|c||r|c|}
\hline
\multicolumn{2}{|c||}{First Regge trajectory} &
\multicolumn{2}{c|}{Second Regge trajectory} \\
\hline
$E\ ({\rm MeV})$ & $\ell$ & $E\ ({\rm MeV})$ & $\ell$ \\
\hline
-4.571182814  & 0 & -0.884280776 & 0 \\
-2.619884138  & 1 & 0.807634844  & $1.000000000+i0.000000034 $   \\
-0.759532418 & 2 & 2.384151637   & $2.000000000+i0.000027431 $   \\
 1.009031953   & $3.000000000+i0.000000013 $ &
 3.810922062   & $2.999963480+i0.003081380 $    \\
 2.676524768   & $4.000000000+i0.000008632 $ &
 5.025176235   & $3.990736977+i0.065066318 $    \\
\hline
\end{tabular}
\end{table}
%
\begin{table}
\caption[]{Third and fourth Regge trajectories for the potential
$V_2(r)\,$. These trajectories start  from the lowest
$S$--wave resonances. }
\begin{tabular}{|r|r||r|r|}
\hline
\multicolumn{2}{|c||}{Third Regge trajectory} &
\multicolumn{2}{c|}{Fourth Regge trajectory} \\
\hline
$E\ ({\rm MeV})$ & $\ell\phantom{+i0.000000000}$ & $E\ ({\rm MeV})$ &
$\ell\phantom{+i0.000000000}$ \\
\hline
2.252380731  & $-0.000000010+i0.000041610 $ &
4.500948186  & $-0.037200122+i0.130360179 $   \\
3.577386775  & $ 0.999888289+i0.005370144 $ &
5.312239776  & $ 0.726629120+i0.451038159 $   \\
4.659668666  & $ 1.976641344+i0.104936778 $ &
6.163323809  & $ 1.389037872+i0.838973703 $   \\
5.581406240  & $ 2.805402233+i0.392410862 $ &
7.091723079  & $ 2.025517176+i1.253031383 $   \\
6.522259887  & $ 3.530415810+i0.760317663 $ &
8.059380065  & $ 2.634772674+i1.671995431 $   \\
\hline
\end{tabular}
\end{table}
\newpage
\def\emline#1#2#3#4#5#6{%
       \put(#1,#2){\special{em:moveto}}
       \put(#4,#5){\special{em:lineto}}}
\def\newpic#1{}
\begin{figure}
\vspace*{3cm}
\centering
\unitlength=1.00mm
\special{em:linewidth 0.4pt}
\linethickness{0.4pt}
\begin{picture}(111.00,100.74)
\put(28.00,95.67){\line(1,0){83.00}}
\put(30.00,97.67){\line(0,-1){42.00}}
\put(30.00,95.67){\line(3,-1){80.00}}
\put(28.04,75.63){\line(1,0){1.96}}
\put(28.04,55.68){\line(1,0){1.96}}
\put(49.95,97.66){\line(0,-1){1.96}}
\put(110.04,97.54){\line(0,-1){1.99}}
\put(90.06,97.66){\line(0,-1){2.11}}
\put(69.95,97.66){\line(0,-1){1.99}}
\put(82.03,94.81){\circle*{1.50}}
\put(93.01,88.66){\circle*{1.50}}
\put(98.06,75.70){\circle*{1.50}}
\put(29.97,100.74){\makebox(0,0)[cb]{0}}
\put(49.96,100.74){\makebox(0,0)[cb]{1}}
\put(69.95,100.74){\makebox(0,0)[cb]{2}}
\put(90.16,100.52){\makebox(0,0)[cb]{3}}
\put(24.92,95.69){\makebox(0,0)[rc]{0}}
\put(24.92,75.48){\makebox(0,0)[rc]{-1}}
\put(18.00,55.72){\makebox(0,0)[lb]{${\rm Im\,}k$}}
\put(109.92,103.20){\makebox(0,0)[rt]{${\rm Re\,}k$}}
\put(49.08,92.17){\makebox(0,0)[lc]{$\theta$}}
\put(67.97,95.03){\circle{1.50}}
\put(87.96,91.73){\circle{1.50}}
\put(94.11,78.77){\circle{1.50}}
\emline{44.91}{90.69}{1}{45.93}{92.29}{2}
\emline{45.93}{92.29}{3}{46.52}{94.00}{4}
\emline{46.52}{94.00}{5}{46.78}{95.65}{6}
\end{picture}
\caption{
The zeros of the Jost function for the $S$--wave resonances
of the potential $V_1(r)$ with (open circles) and without (filled
circles) the Coulomb term. Exact values are given in Table I. The
dividing line corresponds to the rotation angle $\theta=0.1\pi$.}
\label{ttt:pic1}
\end{figure}
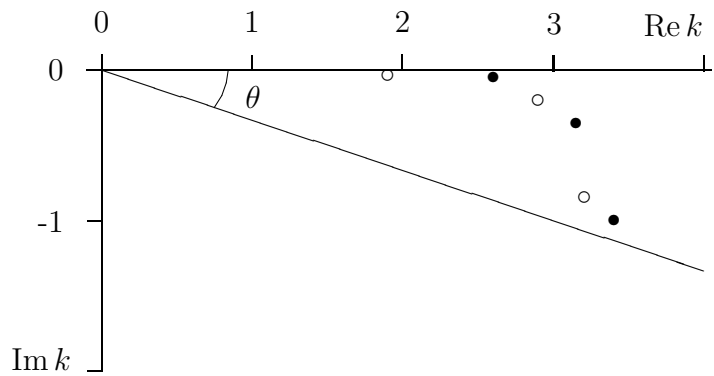
\def\emline#1#2#3#4#5#6{%
       \put(#1,#2){\special{em:moveto}}
       \put(#4,#5){\special{em:lineto}}}
\def\newpic#1{}
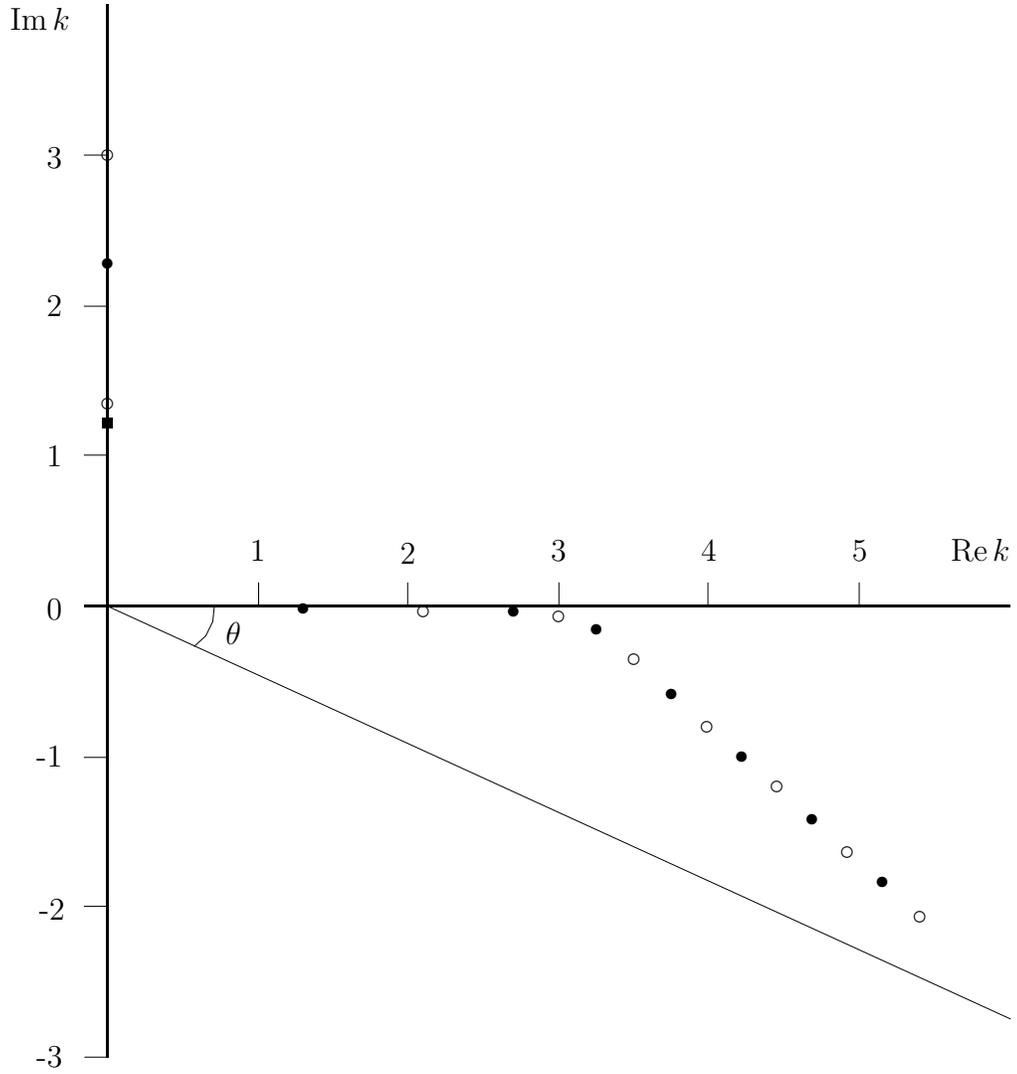
\begin{figure}
\centering
\unitlength=1.00mm
\special{em:linewidth 0.4pt}
\linethickness{0.4pt}
\begin{picture}(140.07,150.00)
\put(17.00,70.00){\line(1,0){123.00}}
\put(20.00,150.00){\line(0,-1){140.00}}
\emline{20.03}{70.00}{1}{140.07}{15.08}{2}
\emline{40.13}{70.00}{3}{40.13}{73.12}{4}
\emline{59.95}{70.00}{5}{59.95}{73.12}{6}
\emline{80.05}{73.12}{7}{80.05}{70.00}{8}
\emline{99.87}{73.12}{9}{99.87}{70.00}{10}
\emline{119.97}{73.12}{11}{119.97}{70.00}{12}
\emline{16.91}{90.11}{13}{20.03}{90.11}{14}
\emline{16.91}{109.92}{15}{20.03}{109.92}{16}
\emline{16.91}{130.03}{17}{20.03}{130.03}{18}
\emline{16.91}{49.90}{19}{20.03}{49.90}{20}
\emline{16.91}{30.08}{21}{20.03}{30.08}{22}
\put(46.00,69.67){\circle*{1.50}}
\put(62.00,69.33){\circle{1.50}}
\put(74.00,69.33){\circle*{1.50}}
\put(80.00,68.67){\circle{1.50}}
\put(90.00,63.00){\circle{1.50}}
\put(99.67,54.00){\circle{1.50}}
\put(109.00,46.00){\circle{1.50}}
\put(128.00,28.67){\circle{1.50}}
\put(118.33,37.33){\circle{1.50}}
\put(85.00,67.00){\circle*{1.50}}
\put(95.00,58.33){\circle*{1.50}}
\put(104.33,50.00){\circle*{1.50}}
\put(113.67,41.67){\circle*{1.50}}
\put(123.00,33.33){\circle*{1.50}}
\put(20.00,130.00){\circle{1.50}}
\put(20.00,97.00){\circle{1.50}}
\put(20.00,115.67){\circle*{1.50}}
\put(19.33,93.67){\rule{1.33\unitlength}{1.33\unitlength}}
\put(40.00,76.00){\makebox(0,0)[cb]{1}}
\put(60.00,75.67){\makebox(0,0)[cb]{2}}
\put(80.00,76.00){\makebox(0,0)[cb]{3}}
\put(100.00,76.00){\makebox(0,0)[cb]{4}}
\put(120.00,76.00){\makebox(0,0)[cb]{5}}
\put(140.00,76.00){\makebox(0,0)[rb]{${\rm Re\,}k$}}
\put(14.00,129.67){\makebox(0,0)[rc]{3}}
\put(14.00,110.00){\makebox(0,0)[rc]{2}}
\put(14.00,90.00){\makebox(0,0)[rc]{1}}
\put(14.00,69.67){\makebox(0,0)[rc]{0}}
\put(14.00,50.00){\makebox(0,0)[rc]{-1}}
\put(14.33,29.67){\makebox(0,0)[rc]{-2}}
\emline{17.00}{10.00}{23}{20.00}{10.00}{24}
\put(14.00,10.00){\makebox(0,0)[rc]{-3}}
\put(15.00,149.67){\makebox(0,0)[rt]{${\rm Im\,}k$}}
\emline{31.56}{64.67}{25}{33.08}{66.07}{26}
\emline{33.08}{66.07}{27}{34.01}{67.94}{28}
\emline{34.01}{67.94}{29}{34.24}{70.04}{30}
\put(35.76,66.42){\makebox(0,0)[lc]{$\theta$}}
\end{picture}
\caption{
The zeros of the Jost function corresponding to bound and resonant
states generated by the potential $V_2(r)$ in the $S$-- (open circles) and
$P$-- (filled circles) partial waves. The filled box indicates the
$D$--wave bound state.  Exact values are given in Table II and III. The
dividing line corresponds to the rotation angle $\theta=0.15\pi$.}
\label{ttt:pic2}
\end{figure}
\def\emline#1#2#3#4#5#6{%
       \put(#1,#2){\special{em:moveto}}
       \put(#4,#5){\special{em:lineto}}}
\def\newpic#1{}
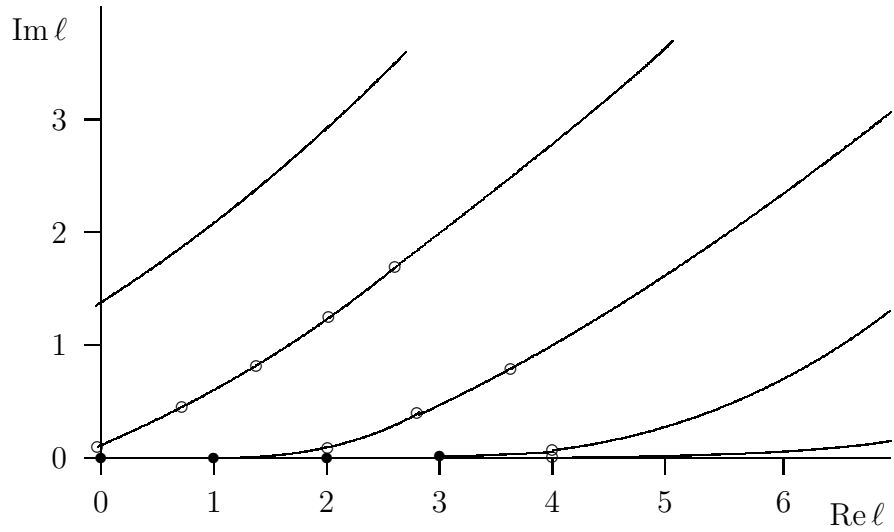
\begin{figure}
\centering
\unitlength=0.75mm
\special{em:linewidth 0.4pt}
\linethickness{0.4pt}
\begin{picture}(150.00,110.00)
\put(7.00,30.00){\line(1,0){143.00}}
\put(10.00,110.00){\line(0,-1){83.00}}
\put(7.00,50.00){\line(1,0){3.00}}
\put(7.00,70.00){\line(1,0){3.00}}
\put(7.00,90.00){\line(1,0){3.00}}
\put(30.00,30.00){\line(0,-1){3.00}}
\put(50.00,27.00){\line(0,1){3.00}}
\put(70.00,27.00){\line(0,1){3.00}}
\put(90.00,27.00){\line(0,1){3.00}}
\put(110.00,27.00){\line(0,1){3.00}}
\put(131.00,27.00){\line(0,1){3.00}}
\put(4.00,106.00){\makebox(0,0)[rc]{${\rm Im\,}\ell$}}
\put(144.00,22.00){\makebox(0,0)[ct]{${\rm Re\,}\ell$}}
\put(10.00,24.00){\makebox(0,0)[ct]{0}}
\put(30.00,24.00){\makebox(0,0)[ct]{1}}
\put(50.00,24.00){\makebox(0,0)[ct]{2}}
\put(70.00,24.00){\makebox(0,0)[ct]{3}}
\put(90.00,24.00){\makebox(0,0)[ct]{4}}
\put(110.00,24.00){\makebox(0,0)[ct]{5}}
\put(131.00,24.00){\makebox(0,0)[ct]{6}}
\put(4.00,30.00){\makebox(0,0)[rc]{0}}
\put(4.00,50.00){\makebox(0,0)[rc]{1}}
\put(4.00,70.00){\makebox(0,0)[rc]{2}}
\put(4.00,90.00){\makebox(0,0)[rc]{3}}
\put(10.00,30.00){\circle*{2.00}}
\put(30.00,30.00){\circle*{2.00}}
\put(50.00,30.00){\circle*{2.00}}
\put(70.00,30.31){\circle*{2.00}}
\put(90.00,30.10){\circle{2.00}}
\put(90.00,31.38){\circle{2.00}}
\bezier{268}(90.00,31.33)(122.67,35.33)(149.67,56.00)
\bezier{80}(70.07,30.31)(74.37,30.13)(89.96,31.02)
\put(50.11,31.84){\circle{2.00}}
\put(65.96,37.92){\circle{2.00}}
\put(82.58,45.78){\circle{2.00}}
\put(9.33,32.00){\circle{2.00}}
\put(24.33,39.00){\circle{2.00}}
\put(50.33,55.00){\circle{2.00}}
\put(62.08,63.85){\circle{2.00}}
\bezier{252}(9.33,32.00)(38.33,44.33)(62.33,64.00)
\put(37.55,46.19){\circle{2.00}}
\bezier{252}(111.33,104.00)(100.00,93.00)(62.33,64.00)
\bezier{288}(9.00,57.00)(37.00,74.67)(64.00,102.00)
\bezier{240}(90.33,30.00)(134.33,30.33)(150.00,33.00)
\bezier{156}(30.00,30.00)(51.67,29.67)(66.33,38.00)
\bezier{400}(66.33,37.67)(102.67,54.33)(150.00,91.33)
\end{picture}
\caption{
 The five lowest Regge trajectories for the potential
$V_2(r)$. Exact values of the points are given in Table IV and V.
Filled circles indicates points which coincide (they are at the same place
of the $\ell$--plane but correspond to different energies) or cannot be
distinguished. }
\label{ttt:pic3}
\end{figure}
\newpage

\end{document}